\documentclass[11pt]{article}
\usepackage{moriond,epsfig}

\def\be{\begin{equation}}
\def\ee{\end{equation}}
\def\bea{\begin{eqnarray}}
\def\eea{\end{eqnarray}}

\begin{document}
\vspace*{4cm}
\title{NEUTRON STARS AS LABORATORIES FOR COSMOLOGY}

\author{ M \'ANGELES P\'EREZ GARC\'IA\footnote{mperezga@usal.es}}

\address{Department of Fundamental Physics and IUFFyM, University of Salamanca,\\ Plaza de la Merced s/n 37008 Salamanca, Spain}

\maketitle\abstracts{Neutron stars can be considered a useful and interesting laboratory for Cosmology. With their deep gravitational potential they may accrete dark matter from the galactic halo and subsequent self-annihilation processes could induce an indirect observable signal this type of matter. In addition, the large densities in the interior of these objects may constitute  a test-bench to study hypothesized deviations of fundamental constant values complementary to existing works using  constraints at low density from BBN. }

\section{Introduction}

Neutron stars (NS) are astrophysical objects where matter is subject to extreme conditions. They were hypothesized on theoretical grounds early in the 1930’s by Baade, Zwicky and Landau and later discovered by Hewish and Bell \cite{hewit}. They are born in the aftermath of a supernova event and typically have a mass less than $M \approx 2 M_\odot$ and a radius $r\approx12$ km. Based on their internal structure the estimated central densities may be about a few times nuclear saturation density $n_0\approx 0.145$  $fm^{-3}$, that is a mass density $\rho\approx 2\, 10^{14}$  $g/cm^3$.  Zero temperature scenarios are usually assumed since in the interior they are in the few keV range while the Fermi energies of the degenerate baryonic species are in the MeVs. Its constituents vary  according to a density distribution from a low density myriad of nuclei to, at larger densities, a hadronic sector composed of neutrons and to a less extent of protons and heavier particles. Leptons keep electrical charge neutrality in the system. In addition to the extreme densities some of them have large magnetic fields with strengths ranging  $B\simeq10^9-10^{15}$ G on the surface. These may power electromagnetic emission that can be detected on Earth, with periods $P \approx 10^{-3}-1$ s due to a misalignement of the rotation and magnetic axis. 

\section{Neutron stars as accretors of dark matter}

Due to their compact size their gravitational potential well $\Phi\approx -GM^2/r$ is able to capture dark matter (DM) from the galactic distribution. Also the sun, other evolved stars and planets may accrete DM. Profiles, $\rho_{DM}$, are not yet fully determined but some of the most popular include work performed by Navarro et al \cite{nfw} or those based on simulations \cite{simu}.  Currently, DM interaction cross sections with matter in the Standard Model are not known and there is a large experimental effort to try to constrain them. Roughly speaking, indirect detection techniques try to obtain the gamma-ray outcome from the self-­annihilation of DM particles while in direct detection they focus on nuclear recoiling from the scattering of the galactic flow of DM as it traverses the Earth \cite{reviewdm}. Additionally, collider searches try to produce and detect SUSY DM candidates by missing transverse energy \cite{collider} in reactions involving Standard Model particles. 
%%%%%%%%%%%%%%%%%%%%%%%%%%%
\begin{figure}[h]
\begin{center}
\epsfig{figure=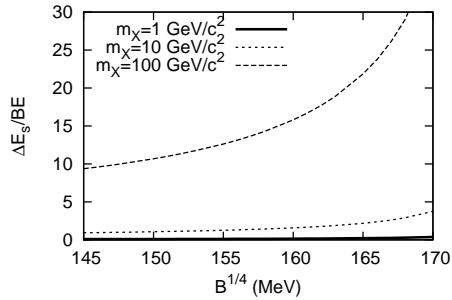,width=0.325\textwidth, angle=-90, scale=0.75}
\end{center}
\caption{Ratio of spark to strangelet binding energy in the nuclear medium as a function of  $B^{1/4}$. \label{fig1}}
\end{figure}
%%%%%%%%%%%%%%%%%%%%%%%%%%%
The  DM-nucleon scattering cross sections are, in principle, mass and energy dependent, but the usual approach in astrophysical scenarios is to assume $S$-wave scattering since at low energy Maxwellian distributed DM particles in a galactic halo will hit the experimental observer setting on Earth with mean velocities $v\simeq270$ km/s$=10^{-3}c$. Considering coherence effects in the cross sections for DM and due to the fact that scaling with the nuclear target chiefly grows with mass number $A$, one can deduce the spin-independent (SI) cross section values  $\sigma_{XN} \simeq 10^{-40}$ $\rm cm^2$ \cite{cdms}. In a NS most of the matter is concentrated in the inner core and estimations for the capture rates of DM particles with mass $m_X$ at the maximum of the galactic plane pulsar distribution at about 3 kpc are given by\cite{perez} $F=\frac{3.042 \,10^{25}}{m_X (GeV)}\frac{\rho_{DM}}{\rho_{DM,0}}\,\,s^{-1}$ where ${\rho_{DM,0}}=0.3$ $\rm GeV/cm^3$ is the DM density at the solar circle. Current indications from DAMA/LIBRA and CoGeNT seem to fit well with a light particle $\approx4-12$ $\rm GeV/c^2$ although other experiments give null results \cite{cdms}. The mean free paths for these particles in NSs can be estimated as $\lambda=1/(\sigma_{XN} n_0) \approx 0.7$ m. The number of in-medium scatterings would be on average $r/\lambda \simeq 1.7\, 10^4$ times constituting an efficient capture.  One of the possible models for DM is that where the particle candidate is of Majorana type and, therefore, self-annihilates with a probability per unit volume $V$ given by $d\Gamma/dV \approx < \sigma v >(\rho_{DM}/m_X)^2$. At sufficiently high density these self-annihilation processes are removing part of the DM from the accreted  distribution that is building up inside the NS. In this approximation the number of particles, $N(t)$, in the interior at time $t$ can be obtained solving the equation $\frac{dN(t)}{dt}\approx F -\Gamma$.
Let us consider that the energy deposit in the nuclear medium could be producing a series of {\it sparks} of the order of a fraction $f$ of the $X$ mass in the few GeV range as $\Delta E_{spark}=2 f m_X c^2$. This seeding mechanism may produce one or multiple sparks that may help partially deconfine  the hadronic quark content at large densities in the center of those objects \cite{perez}. These aggregates of quarks are known as {\it strangelets} and are currently under search in the CASTOR calorimeters at the CMS experiment in the LHC or AMS in the ISS. The binding energy (BE) of such a strangelet can be calculated and it is given in the MIT model by\cite{madsen}  $E_A(\mu_i,m_i,B)+ E_{Coul}$ where $\mu_i$ and $m_i$ are the chemical potential and mass of the $ith$-type quark, respectively. B is the MIT bag constant and $\rm E_{Coul}$ is the correction due to electrical charge. In Fig.{\ref{fig1}} we show the ratio of spark to strangelet binding energy with baryon number $A= 10$ in the nuclear medium for $m_X =1, 10, 100$ $\rm GeV/c^2$ as a function of $B^{1/4}$ for $f =0.1$ at a central density of $2n_0$ compatible with NS observations. Even in this very conservative case where only $10\%$ on the energy is deposited in the medium may indeed be larger than the strangelet binding. 
%%%%%%%%%%%%%%%%%%%%%%%%%%%
\begin{figure}[h]
\begin{center}
\epsfig{figure=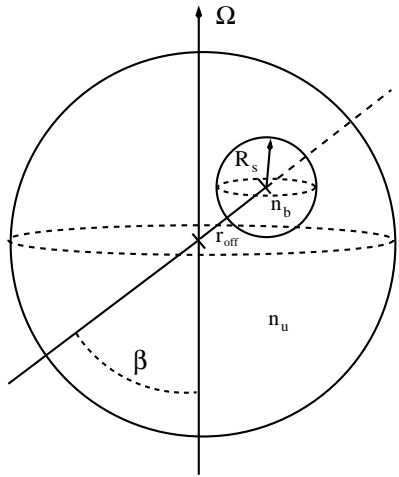,width=0.325\textwidth, scale=0.6}
\caption{Simplified scheme of one bubble of quark matter proceeding through NS matter. \label{fig2}}
\end{center}
\end{figure}
%%%%%%%%%%%%%%%%%%%%%%%%%%%
Thus, DM seeding may act as a 'Trojan Horse' liberating energy inside the central regions and cause a dramatic change in the internal structure of the object.The off-center energy release may have a kinematical signature from a velocity kick and change in the rotation pattern \cite{perez}. More exotic possibilities include  the LSP in the SUSY theories as the DM candidate. If they are accelerated enough towards a gravitational compact object like a black hole or NS they could annihilate to an slepton and then radiate out a SM lepton and another LSP, provided the mass difference between the NSLP and LSP are up to a few GeV \cite{mirco}.

\subsection{Kinematical implications of DM seeding in NS}

Once the seeding mechanism has taken place at distance $r_{off}$ from the NS center, then the quark bubbles (with density $n_b$ and radius $R_s$) in the hadronic medium of density $n_u$, may coalescence\cite{perez}. In this way the progression of a burning front may proceed in direction $\beta$, for a simplified scheme of this see Fig. \ref{fig2}. NS burning has been partially studied \cite{burn} although the physical mechanisms for this are not yet clear and the hadron to quark star (QS) transition was only hypothesized in the context of very large central densities or temperature fluctuations. Obser­vationally, the signature for this transition could be the emission of a gamma-ray-burst (GRB) due to the change in the gravitational and baryonic energy density. Both are of the same order and depend on the radial structure change $\Delta E_G\approx  \frac{\Delta r}{r}10^{53} \, \rm erg$. This is in agreement with order of magnitude of some of the Fermi measurements but further analysis is needed.  If certain, this induced NS to QS conversion could provide an {\it internal engine} in these events where NS crust ejection is supposed to happen.

\section{Testing variations of the fundamental couplings in extreme conditions}

The extreme pressure and energy density attained in the interior of NSs can be also used to test some models in grand unification theories (GUT). Following GUT prescription \cite{coc} the set of fundamental forces: electroweak, strong and gravitational could be unified at some unspecified large energy scale. If we consider that the electromagnetic cou­pling constant $\alpha$ may vary with density contrasts, one could use some low density constraints from BBN \cite{bbn} and, accordingly, NSs seem to be a good test-bench for trying to understand the large density counterpart.
Using current information from either terrestrial \cite{dani}(Heavy ion collisions, low density simulations of pure neutron matter, etc ) and astrophysical sources \cite{steiner} the equation of state (EOS) of nuclear matter constrains possible variations of the couplings that are not consistent with supporting the interplay of these three forces in a realistic NS. Using a physically motivated relativistic mean field lagrangian model parametrization, combined variation of  couplings can be studied \cite{martins} since the expres­sions of pressure and energy density in the EOS consistently modify. The variations in $\Delta \alpha/\alpha$ relate to those in the gravitational constant $G$ and meson masses for the strong interaction. We consider an input spatial variation in the allowed range by CMB constraints \cite{mene}, $\Delta \alpha/\alpha \simeq \pm 10^{-2}$. We take for example $\Delta \alpha/\alpha=0.005$ and what we find \cite{martins} is that the corresponding allowed regions, from an initial square $L^2$ centered at origin and side $L=900$, for $R, S$ parameters lie in the triangular shape between,
\be
S\approx450-3.8(R+100);\,\,\, S\approx450-0.145\,R.
\ee
The relative variations for particle masses are \cite{coc} $\frac{\Delta m_e}{m_e}= 0.5(1+S)\frac{\Delta \alpha}{\alpha}\,$, $\frac{\Delta m_p}{m_p}=  \left[ 0.8R+0.2(1+S)\right] \frac{\Delta \alpha}{\alpha}$, $\frac{\Delta m_n}{m_n} = \left[ (0.1+0.7S-0.6R)+\frac{m_p}{m_n}(0.1-0.5S+1.4R) \right] \frac{\Delta \alpha}{\alpha}$. In a 'canonical' case where $R=20$, $S=160$  and imposing isospin symmetry we have $\frac{\Delta m_i}{m_i}=(0.201, 0.201, 0.5025)$ for $i=p, n, e$. Further work is needed to check for EOS model dependence. However we do not expect a change the general trend obtained. So fr the only constrains that exist  come from a low density environment and our approach is, therefore, complementary.
In this contribution we have discussed on the role of NSs as laboratories for  extreme conditions of matter where some of the most interesting current issues in Cosmology can be tested to provide complementary insight to other existing constraints.
\section*{Acknowledgments}
We would like to acknowledge support under ESF-COMPSTAR and  MICINN Consolider MULTIDARK and FIS-2009-07238 projects.

\end{document}